# Magnified Damping under Rashba Spin Orbit Coupling

Seng Ghee Tan[† 1,2] , Mansoor B. A. Jalil[1,2]

(1) Data Storage Institute, Agency for Science, Technology and Research (A*STAR)
  2 Fusionopolis Way, #08-01 DSI, Innovis, Singapore 138634

(2) Department of Electrical Engineering, National University of Singapore,
  4 Engineering Drive 3, Singapore 117576

**Abstract**

The spin orbit coupling spin torque consists of the field-like **[REF: S.G. Tan et al., arXiv:0705.3502, (2007).]** and the damping-like terms **[REF: H. Kurebayashi et al., Nature Nanotechnology 9, 211 (2014).]** that have been widely studied for applications in magnetic memory. We focus, in this article, not on the spin orbit effect producing the above spin torques, but on its magnifying the damping constant of all field like spin torques. As first order precession leads to second order damping, the Rashba constant is naturally co-opted, producing a magnified field-like damping effect. The Landau-Liftshitz-Gilbert equations are written separately for the local magnetization and the itinerant spin, allowing the progression of magnetization to be self-consistently locked to the spin.

## 1. Introduction

In spintronic and magnetic physics, magnetization switching and spin torque [1] have been well-studied. The advent of the Rashba spin-orbit coupling (RSOC) [2,3] due to inversion asymmetry at the interface of the ferromagnetic/heavy atom (FM/HA) heterostructure introduces new spin torque to the FM magnetization. The field-like [4-6] and the damping-like [7] SOC spin torque had been theoretically derived based on the gauge physics and the Pancharatnam-Berry's phase, as well as experimentally verified and resolved. The numerous observations of spin-orbit generation of spin torque [8-10], are all related to the experimental resolutions [6,7] of their field-like and damping-like nature, thus ushering in the possibility of spin-orbit based magnetic memory. While the damping-like spin torque due to Kurebayashi et al. [7] is dissipative in nature, the field-like due to Tan et al. [4,11], is non-dissipative, and precession causing. Recent studies have even more clearly demonstrated the physics and application promises of both the field-like and the damping-like SOC spin torque [12-14]. Besides, similar SOC spin torque have also been studied theoretically in FM/3D-Rashba [15] and FM-topological-insulator material [16,17], and experimentally shown [18, 19] in topological insulator materials.

The dissipative physics of all field-like magnetic torque terms have been derived in second-order manifestation in a manner introduced by Gilbert in the 1950's. Conventional study of magnetization dynamics is based on a Gilbert damping constant which is incorporated manually into the Landau-Lifshitz-Gilbert (LLG) equation. In this paper, we will focus our attention not so much on the spin-orbit effect producing the SOC spin torque, as on the spin-orbit effect magnifying the damping constant of all field-like spin torques. As field-like spin torques, regardless of origins, generate first-order precession, the Rashba constant will be co-opted into the second-order damping effect, producing a magnified damping constant. On the other hand, conventional incorporation of the dissipative damping physics into the LLG would fail to account for the spin-orbit magnification of the damping strength. It would therefore be necessary to derive the LLG equations from a Hamiltonian which describes electron due to the local FM magnetization($m$), and those itinerant ($s$) and injected from external parts. We present a set of modified LLG equations for the $m$ and the $s$. This will be necessary for a more precise modeling of the $m$ trajectory that simultaneously tracks the $s$ trajectory. In summary, the two central themes of this paper is our presentation of a self-consistent set of LLG equations under the Rashba SOC and the derivation of the Rashba-magnified damping constant in the second-order damping-like spin torque.

## 2. Theory of Magnified Damping

The system under consideration is a FM/HA hetero-structure with inversion asymmetry provided by the interface. Free electron denoted by $s$, is injected in an in-plane manner into the device. The FM equilibrium electron is denoted by $m$. One considers the external source-drain bias to inject electron of free-electron nature $s$ into the FM with kinetic, scattering, magnetic, and spin-orbit energies. The Hamiltonian is

$$H_f = \frac{p^2}{2m} + V_{imp}^s + J_{sd} S.M + \mu_0 M.H_{ani} + \left(\frac{2(\lambda + \lambda')}{\hbar}\right)(s+m).(p \times E_t)$$
$$- i(\lambda + \lambda')(s+m).(\nabla \times E_t)$$

(1)

where $s, m$ have the units of angular momentum i.e. $n\frac{\hbar}{2}$, while $M = \left(\frac{g_s \mu_B}{\hbar}\right) m$ has the unit of magnetic moment, and $\mu_B = \frac{e\hbar}{2m}$ is the Bohr magneton. Note that $\left(\frac{2\lambda}{\hbar}\right)$ is the vacuum SOC constant, while $\left(\frac{2\lambda'}{\hbar} = \frac{2\eta_R}{\hbar^2 E_{inv}}\right)$ is the Rashba SOC constant. The SOC part of the Hamiltonian illustrates the simultaneous presence of vacuum and Rashba SOC. The proportion of the number of electron subject to each coupling would depend on the degree of hybridization. But since $\lambda' \gg \lambda$, the above can be written with just the Rashba SOC effect. Care is taken to ensure $\lambda, \lambda'$ share the same dimension of $Tesla^{-1}$, and $E_t$ is the total electric field, $J_{sd}$ is the s-d coupling constant, $V_{imp}^s$ denotes the spin flip scattering potential, $H_{ani}$ denotes the anisotropy field of the FM material. On the other hand, one needs to be aware that the above is an expanded SOC expression that comprises a momentum part as well as a curvature part [20]. One can then consider the physics of the electric curvature as related to the time dynamic of the spin moment, which bears a similar origin to the Faraday effect. In the modern context of Rashba physics [21], one considers electron spin to lock to the orbital angular momentum $L$ due to intrinsic spin orbit coupling at the atomic level. Due to broken inversion symmetry, electric field ($E_{inv}$) points perpendicular to the plane of the FM/HA host. Because of hybridization, the $s, L, p$ of an electron is coupled in a complicated way by the electric field. In a simple way, one first considers $L$ to be coupled as $H = \left(\frac{2\lambda}{\hbar}\right) L \cdot (p \times E_{inv})$. As spin $s$ is coupled via atomic spin orbit locking to $L$, an effective coupling of $s$ to $E_{inv}$ can be expected to occur with strength as determined by the atomic electric field. We will now take things a step further to make an assumption that $s$ is also coupled via $L$ to other sources of electric fields e.g. those arising from spin dynamic $\left(\frac{dM}{dt}, \frac{dS}{dt}\right)$, in the same way that it is coupled to $E_{inv}$. The actual extent of coupling will, however, be an experimental parameter that measures the efficiency of Rashba coupling to $E_{inv}$ as opposed to electric fields ($E_m, E_s$) arising due to spin dynamic. The total electric field in the system is now $E_t = E_{inv} + E_m + E_s$, where $E_m, E_s$ arise due to $\frac{dM}{dt}, \frac{dS}{dt}$, respectively. On the momentum part of the Hamiltonian $2\lambda' s \cdot (k \times E_t)$, we only need to consider that $E_t = E_{inv}$ as one can, for simplicity, consider $E_m$ and $E_s$ to simply vanish on average. Thus in this renewed treatment, the momentum part is:

$$\frac{2\lambda'}{\hbar} s \cdot (p \times E_{inv}) = \eta_R \sigma \cdot (k \times e_{inv})$$

(2)

where $\eta_R = \lambda' \hbar E_{inv}$ is the Rashba constant that has been vastly measured in many material systems with experimental values ranging from 0.1 to 2 $eV\text{Å}$. On the curvature part, one considers $E_t = E_m + E_s$ without the $E_{inv}$ as $E_{inv}$ is spatially uniform and thus would have zero curvature. In summary, the theory of this paper has it that the time-dynamic of the spin in a Rashba system produces a curvature part of $i\lambda'(s + m) \cdot (\nabla \times E_t)$. Without the Rashba effect, this energy term would just take on the vacuum constant of $\left(\frac{2\lambda}{\hbar}\right)$ instead of the magnified $\left(\frac{2\lambda'}{\hbar}\right)$. The key physics is that in a Rashba FM/HA system, curvature $i\lambda'(s + m) \cdot (\nabla \times E_t)$ is satisfied by the first-order precession due to $\frac{dM}{dt}, \frac{dS}{dt}$ which provide the electric field curvature in the form of $-\mu_0(1 + \chi_m)\frac{dM}{dt} = \nabla \times E_m$, and $-\mu_0(1 + \chi_s)\frac{dS}{dt} = \nabla \times E_s$, where we remind reader again that $M, S$ have the unit of magnetic moment. This results in spin becoming coupled to its own time dynamic, producing a spin-orbit second-order damping-like spin torque. The electric field effect is illustrated in Fig.1 below:

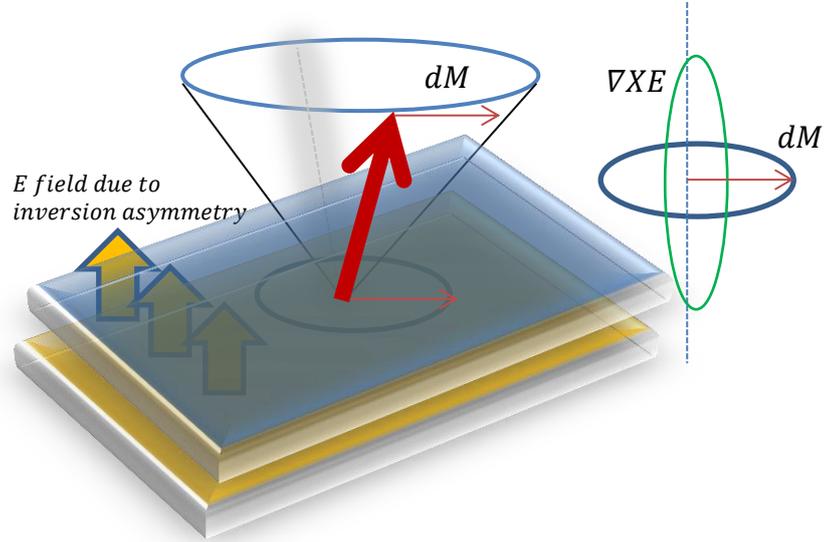

Fig.1. Magnetic precession under the effect of electric fields due to inversion asymmetry, self-dynamic of $\frac{d\mathbf{M}}{dt}$ and the spin dynamic of $\frac{d\mathbf{S}}{dt}$. Projecting $dM$ to the heterostructure surface, one could visualize the emergence of an induced electric field in the form of $\nabla \times E$ in such orientation as to satisfy the law of electromagnetism.

One notes that the LLG equation is normally derived by letting $\mathbf{S}$ satisfy the physical requirements of spin transport. One example of these requirements is assumed and discussed in **REF 1**, with definitions contained therein:

$$\mathbf{S}(r,t) = S_0 \mathbf{n} + \boldsymbol{\delta S}$$

$$\mathbf{J}(r,t) = \frac{-\mu_B P}{e} \mathbf{J}_e \otimes \mathbf{n} - D_0 \nabla \boldsymbol{\delta S}$$

(3)

where $\mathbf{n}$ is the unit vector of $\mathbf{M}$, and $D_0$ is the spin diffusion constant. Thus $\mathbf{S} = \mathbf{S_0} + \boldsymbol{\delta S}$ would be the total spin density that contains, respectively, the equilibrium, the non-equilibrium adiabatic, non-adiabatic, and Rashba field-like terms, i.e. $\boldsymbol{\delta S} = \boldsymbol{\delta S}_a + \boldsymbol{\delta S}_{na} + \boldsymbol{\delta S}_R$. One notes that $\mathbf{S_0}$ is the equilibrium part of $\mathbf{s}$ that is aligned to $\mathbf{m}$, meaning $\mathbf{s_0}$ could exist in the absence of external field and current in the system. The conditions to satisfy are represented explicitly by the equations of:

$$\frac{\partial \delta S}{\partial t} = 0, \quad D_0 \nabla^2 \delta S = 0, \quad \frac{-\mu_B P}{e} \nabla \cdot J_e \frac{M}{M_s} = 0, \quad s_0 \frac{M(r,t)}{t_f M_s} = 0$$

(4)

In the steady state treatment where $\frac{\partial \, \delta S}{\partial t} = 0$, one recovers the adiabatic component of $\delta S_a = n \times j_e . \nabla n$, and the non-adiabatic component of $\delta S_{na} = j_e . \nabla n$. We also take the opportunity here to reconcile this with the gauge physics of spin torque, in which case, the spin potential $A_\mu^{sm} = e \left[ \alpha \, UE_i \sigma_j \varepsilon_{ij\mu} U^\dagger + \frac{i\hbar}{e} U \partial_\mu U^\dagger \right]$ would correspond, respectively, to $\delta S_R + \delta S_a$. In fact, the emergent spin potential **[22, 23]** can be considered to encapsulate the physics of electron interaction with the local magnetization under the effect of SOC **[4, 5, 24-26]**. Here we caution that $\delta s_R$ is restricted to the field-like spin-orbit effect only.

However, in this paper, $S$ is defined to satisfy the transport equations in Eq.(4) except for $\frac{\partial \delta S}{\partial t} = 0$. Keeping the dynamic property of $S$ here allows a self-consistent equation set $\frac{dS}{dt}, \frac{dM}{dt}$ to be introduced. The energy as experienced by the $S, M$ electron are, respectively,

$$H_{fs} = S \cdot \frac{\delta H_f}{\delta S}, \quad H_{fm} = M \cdot \frac{\delta H_f}{\delta M}$$

(5)

with caution that $H_{fs} \neq H_{fm}$. Upon rearrangement, the $s, m$ centric energies are, respectively,

$$H_{fs} = \left( \frac{p^2}{2m} + V_{imp}^S + J_{sd} S \cdot M + S \cdot B_R - i\lambda' s \cdot (\nabla \times E_t) \right)$$

$$H_{fm} = (J_{sd} M \cdot S + \mu_0 M \cdot H_a - i\lambda' m \cdot (\nabla \times E_t))$$

(6)

where $\frac{2\lambda'}{\hbar} s \cdot (p \times E_t) = S \cdot B_R$, while $\frac{2\lambda'}{\hbar} m \cdot (p \times E_t)$ vanishes. We particularly note that there have been recent discussions on the field-like **[4,6,11]** spin orbit torque as well as the damping **[7]** version. With $\frac{ds}{dt} = -\frac{1}{i\hbar}[s, H_{fs}], \frac{dm}{dt} = -\frac{1}{i\hbar}[m, H_{fm}]$, one would now have four dissipative torque terms experienced by electron $s, m$ as shown below:

$$\begin{pmatrix} \tau_{SS} & \tau_{SM} \\ \tau_{MS} & \tau_{MM} \end{pmatrix} = i\lambda' \mu_0 \begin{pmatrix} s \times (1 + \chi_s^{-1})\frac{dS}{dt} & s \times (1 + \chi_m^{-1})\frac{dM}{dt} \\ m \times (1 + \chi_s^{-1})\frac{dS}{dt} & m \times (1 + \chi_m^{-1})\frac{dM}{dt} \end{pmatrix}$$

(7)

To be consistent with conventional necessity to preserve magnetization norm in the physics of the LLG equation, we will drop the off-diagonal terms which are norm-breaking (non-conservation). This is in order to keep the LLG equation in its conventional norm-conserving form, simplifying physics and calculation therefrom. Nonetheless, the non-conserving parts represent new dynamic physics that can be analysed in the future with techniques other than the familiar LLG equations. The self-consistent pair of spin torque equations in their open forms are:

$$\frac{\partial S}{\partial t} = -\left(S \times B_R + \frac{S}{t_f}\right) - \frac{1}{e}\nabla_a(j_a^s S) - \left(\frac{S \times M}{m t_{ex}}\right) - \tau_{SS}$$

$$\frac{\partial M}{\partial t} = -\gamma M \times \mu_0 H_a - M \times \frac{S}{m\, t_{ex}} - \tau_{MM}$$

(8)

where $J_{sd} = \frac{1}{m t_{ex}}$ has been applied, $\gamma$ is the gyromagnetic ratio, $\chi_m$ is the susceptibility. For the study of Rashba-magnified damping in this paper, we only need to keep the most relevant term which is $\tau_{MM} = \frac{i\,\eta_R}{\hbar E_{inv}} \mu_0 (1 + \chi_m^{-1})\, m \times \frac{d\,M}{dt}$. In the phenomenological physics of Gilbert, the first-order precession leads inevitably to the second-order dissipative terms via $s.\frac{dS}{dt}$, $m.\frac{dM}{dt}$. But in this paper, the general SOC physics had been expanded as shown in earlier sections, so that the dissipative terms are to naturally arise from such expansion. The advantage of the non-phenomenological approach is that, as said earlier, the Rashba constant will be co-opted into the second-order damping effect, resulting in the magnification of the damping constant associated with all field-like spin torque.

## 3. Conclusion

The important result in this paper is that the damping constants have been magnified by the Rashba effect. This would not be possible if the damping constant was incorporated manually by standard means of Gilbert. As the Rashba constant is larger than the vacuum SOC constant as can be deduced from Table 1 and shown below

$$\alpha_R = \alpha \frac{\lambda'}{\lambda},$$

(9)

magnetization dynamics in FM/HA hetero-structure with inversion asymmetry (interface, or bulk) might have to be modelled with the new equations. It is important to remind that all previously measured $\eta_R$ has had $E_{inv}$ captured in the measured value. But what is needed in our study is the coupling of $S$ to a dynamic electric field, and that requires the value of just the coupling strength ($\lambda'$). As most measurement is carried out for $\eta_R$, the exact knowledge of $E_{inv}$ corresponding to a specific $\eta_R$ will have a direct impact on the actual value of $\lambda'$. We will, nonetheless, provide a quick, possibly exaggerated estimate. Noting that $\lambda = \frac{e\hbar}{4m^2c^2}$ and $\lambda' = \frac{\eta_R}{\hbar E_{inv}}$, and taking one measured value of $\eta_R = 1 \times 10^{-10} eVm$, corresponding to a $E_{inv} = 10^{10} V/m$, the magnification of $\alpha$ works out to $10^4$ times in magnitude, which may seem unrealistically strong. The caveat lies in the exact correspondence of $\eta_R$ to $E_{inv}$, which remains to be determined experimentally. For example, if an experimentally determined $\eta_R$ actually corresponds to a much larger $E_{inv}$, that would mean that $\lambda' = \frac{\eta_R}{\hbar E_{inv}}$ which magnifies the damping constant through $\alpha_R = \alpha \frac{\lambda'}{\lambda}$ might actually be much lower than present estimate. Therefore, it is worth remembering, for simplicity sake that $\alpha_R$ actually depends on the ratio of $\frac{\eta_R}{E_{inv}}$ but not $\eta_R$. It has also been assumed that $L$ couples to $E_s, E_m$ with the same efficiency that it couples to $E_{inv}$. This is still uncertain as the Rashba constant with respect to $E_s, E_m$ might actually be lower than those $\eta_R$ values that have been experimentally measured mostly with respect to $E_{inv}$. Last, we note that as damping constant has been magnified here, and as increasingly high-precision, live monitoring of simultaneous $s, m$ evolution is no longer redundant in smaller devices, care has been taken to present the LLG equations in the form of a self-consistent pair of dynamic equations involving $M$ and $S$. This will be necessary for the accurate modeling of the simultaneous trajectory of both $M$ and $S$.

Table 1. Summary of damping torque and damping constant with and without Rashba effects.

| | Hamiltonian | Torque | Damping constant |
|---|---|---|---|
| 1. | $H = \left(\frac{2\lambda}{\hbar}\right) s \cdot (p \times E_t)$ <br> $\lambda = \frac{e\hbar}{4m^2c^2}$ | $\frac{\partial m}{\partial t} = i\lambda\mu_0 m \times (1 + \chi_m^{-1})\frac{\partial M}{\partial t}$ | $\alpha = \frac{i\lambda}{2}\mu_0 M_s(1 + \chi_m^{-1})$ |
| 2. | $H_R = \left(\frac{2\lambda'}{\hbar}\right) s \cdot (p \times E_t)$ <br> $\lambda' = \frac{\eta_R}{\hbar E_{inv}}$ | $\frac{\partial m}{\partial t} = i\lambda'\mu_0 m \times (1 + \chi_m^{-1})\frac{dM}{dt}$ | $\alpha^R = \frac{i\lambda'}{2}\mu_0 M_s(1 + \chi_m^{-1})$ |
| | | | |